\documentclass[preprint,aps]{revtex4}
\newcommand{\gra}[1]{\mbox{\boldmath $ #1 $}}
\newcommand{\etal}{{\it et al.}}
\input epsf 
\usepackage{graphicx}  

\usepackage{graphicx}
\usepackage{dcolumn}
\usepackage{bm}

\begin{document}

\preprint{APS/123-QED}

\title{One-point statistics and intermittency of induced electric field in the solar wind}

\author{Luca Sorriso-Valvo, Vincenzo Carbone}
 \affiliation{Dipartimento di Fisica, Universit\`a della Calabria and Istituto Nazionale per la 
Fisica della Materia, Sezione di Cosenza, Rende (CS) - Italy.}
 \altaffiliation[Also at ]{Dipartimento di Fisica, Universit\`a della Calabria.}
\author{Roberto Bruno}
\affiliation{Istituto di Fisica dello Spazio Interplanetario, CNR, 00133 Roma, Italy}

\date{\today}

\begin{abstract}
The interplanetary induced electric field $\gra e =\gra v \times \gra b$ is studied, 
using solar wind time series. The probability distribution functions (PDFs) of the electric field 
components are measured from the data and their non-gaussianity is discussed.
Moreover, for the first time we show that the electric field turbulence is characterized by intermittency. 
This point is addressed by studying, as usual, the scaling of the PDFs of field increments, which allows
a quantitative characterization of intermittency.
\end{abstract}

\maketitle

\section{Introduction}

Solar wind provides a great opportunity to get ``in situ" observations of a magnetized
plasma on a wide range of scales. In fact, many spacecrafts have been launched in
the interplanetary space during past years, in order to study the properties of
plasma. The low-frequency turbulence of the wind can be described in the 
framework of the magnetohydrodynamics (MHD) model of plasma (see for example~\cite{frisch,biskamp} 
for ordinary fluid  and MHD turbulence). Turbulent high amplitude fluctuations have been observed from 
data, and it is well known now that solar wind plasma is in a highly turbulent 
state~\cite{burlaga,tuemarsch}. 
Studies of solar wind turbulence have been mainly focused on the analysis of velocity and 
magnetic field data, for example observing the power spectra of such fields~\cite{matthaeus}, 
and later studying the anomalous statistics of the fields fluctuations, which revealed intermittency 
(\cite{burlaga,marschetu,carbone95,sorriso99}). 
Only recently attention has been paid to the measurements 
of the interplanetary induced electric field (IEF) $\gra e =\gra v \times \gra b$ ($\gra v$ and $\gra b$ 
being respectively the measured velocity and magnetic field). Breech \etal~\cite{breech} 
reported on the statistics of IEF collected from many different spacecrafts spanning 30 years of 
measurements. They analysed the one-point probability distribution functions (PDFs) of the IEF 
fluctuations $\gra{e^\prime}=\gra{v^\prime}\times \gra{b^\prime}$ (computed as a Reynolds decomposition, 
{\it e. g.} $\gra{v}=\gra{V_0}+\gra{v^\prime}$), and they found that such PDFs have exponential tails. 
This has been interpreted through a recent analytical result~\cite{milano}. These authors showed that, 
a field that can be written as $\phi=\xi_1\xi_2-\xi_3\xi_4$, where $\xi_1$, $\xi_2$, $\xi_3$, $\xi_4$ 
are independent stochastic variables with gaussian PDF, has exponential PDF. 
The cases of ``dynamo-type'' and ``cross helicity-type'' correlated gaussian PDFs for the $\xi$'s are 
also shown to give a modified exponential PDF for $\phi$. 
Since the interplanetary velocity and magnetic field components PDFs are roughly 
gaussian, results in~\cite{breech} suggest that the data analysed by these authors satisfy the
hypotheses on correlation required in~\cite{milano}. 

In this letter we present a somewhat different data analysis, by considering the main 
statistical properties of the IEF $\gra e$ itself as computed from a more homogeneous dataset. 
At a variance with the IEF fluctuations, we will see how correlations play a key role in invalidating 
the assumptions for the above results to hold, despite the fact that, from the same dataset, 
the same results as in~\cite{breech} are retrived for the IEF fluctuations $\gra{e^\prime}$. 
We then investigate for the first time the intermittency of interplanetary IEF, which we expect to owe
similar properties as for the interplanetary velocity and magnetic field. We do that through the analysis 
of the scaling laws of field fluctuations across different scales, which gives informations about the 
nonlinear processes undergoing the turbulent cascades of the ideal invariants of the flow. Our results
give a quantitative characterization of intermittency, which was not yet present in the literature.

\section{The interplanetary induced electric field}

In order to study the statistical properties of the solar wind induced electric
field, we use the velocity and magnetic fields, as measured {\it in situ} 
by the spacecraft Helios~2 (1976). Helios~2 orbit lied in the ecliptic plane, 
so that the measured data include both fast and slow wind streams. 
Since the physical conditions are very different, we analyse separately six fast 
and five slow wind streams~\cite{tuemarsch,sorriso99}. 
This separation lead to more homogeneous data sets,  although the radial evolution is not
taken into account here.
After separating the streams, and rejecting the velocity boundary regions between streams, 
which include undesired shear effects, the original dataset reduces to two sets of about $10^5$ 
points each. The sampling time is $81$~seconds, and each stream consists of 2187 points, 
so that we can investigate scaling in a range from about one minute up to one day.
First of all, we compute the IEF as $\gra e(t)=\gra v(t)\times\gra b(t)$, where 
$\gra v(t)$ is the measured velocity field, and $\gra b(t)$ is the measured 
magnetic field.
The choice of the reference frame is not trivial. In fact, when analysing
the velocity field the ``natural'' reference frame is the SE (with the $x$ axis along the sun-earth 
direction, which coincides with the mean velocity). On the other hand, the ``natural'' frame for 
the magnetic field is with the $x$ axis along the mean field line, which is directed like the 
Parker's spiral, and evolves with the heliocentric distance. Note that $z$ axis lies on the ecliptic plane,
while $z$ axis is choosen as normal to the ecliptic plane.
Since we have no arguments to decide whether a frame is better than the other in describing
turbulent effects when analysing the induced electric field, we perform the analysis using both frames, 
and we compare the results.
\begin{figure}
   \centerline{
            \hbox{\epsfxsize=6.0 cm \epsfbox{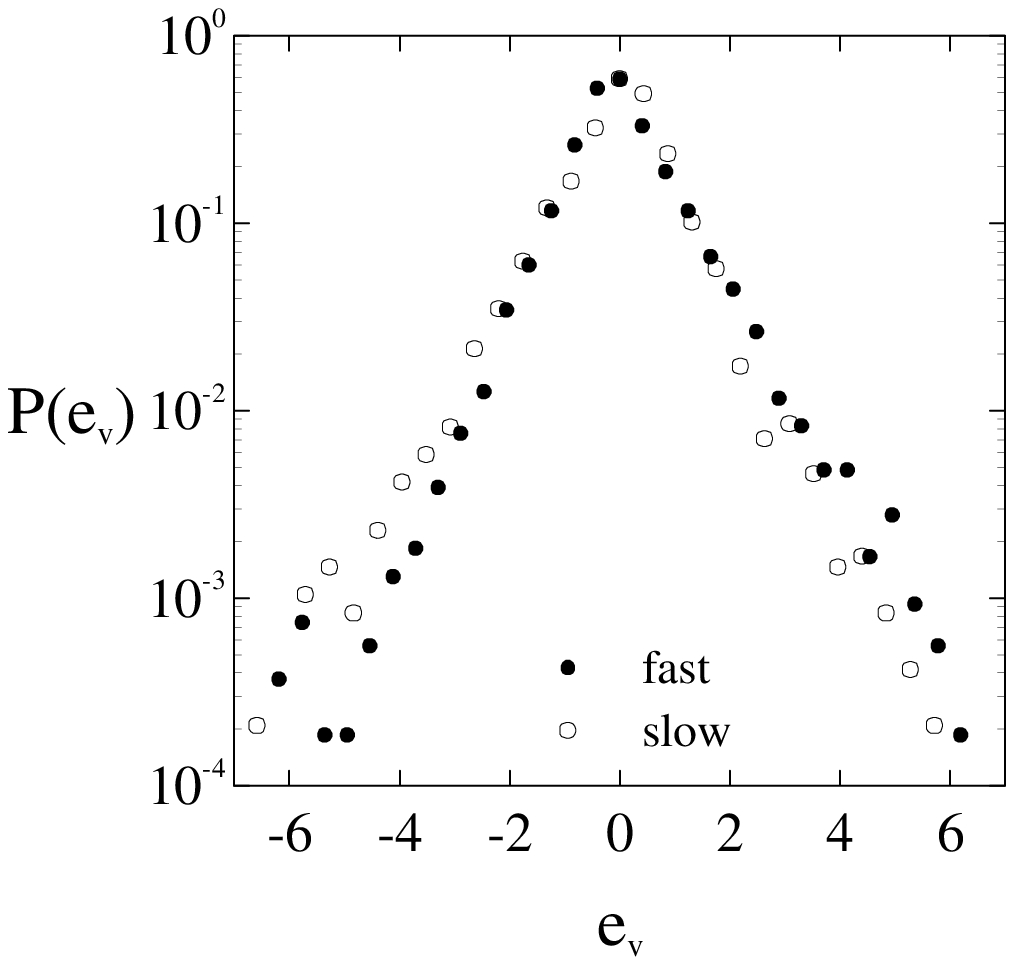}
                     \epsfxsize=6.0 cm \epsfbox{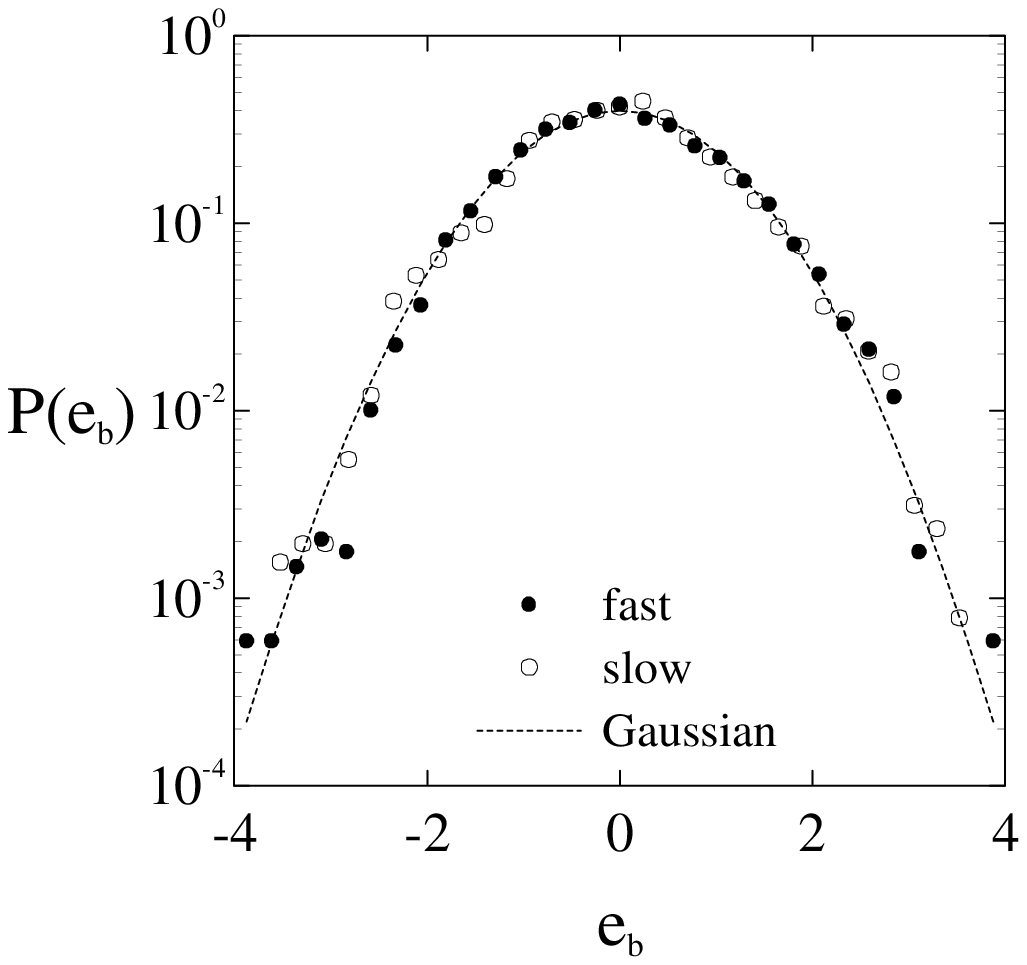}}  
                  }
   \centerline{
            \hbox{\epsfxsize=6.0 cm \epsfbox{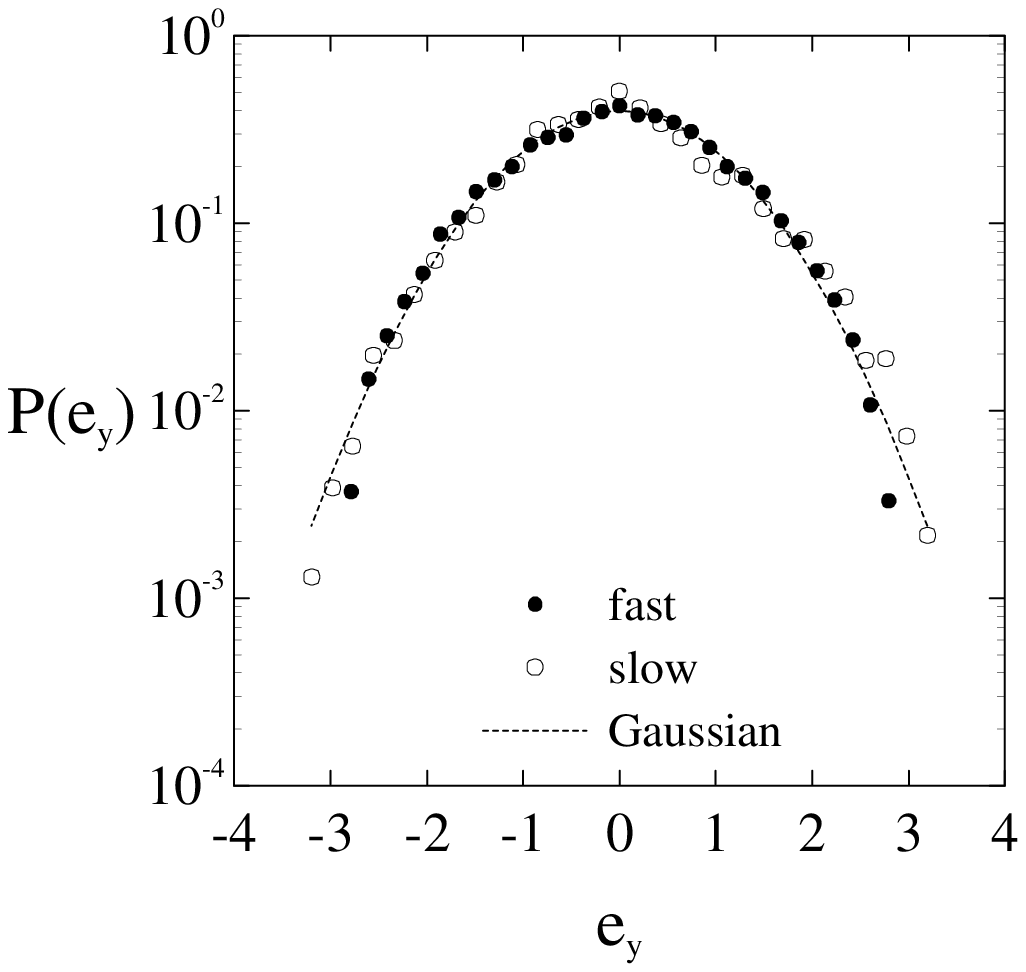}
                     \epsfxsize=6.0 cm \epsfbox{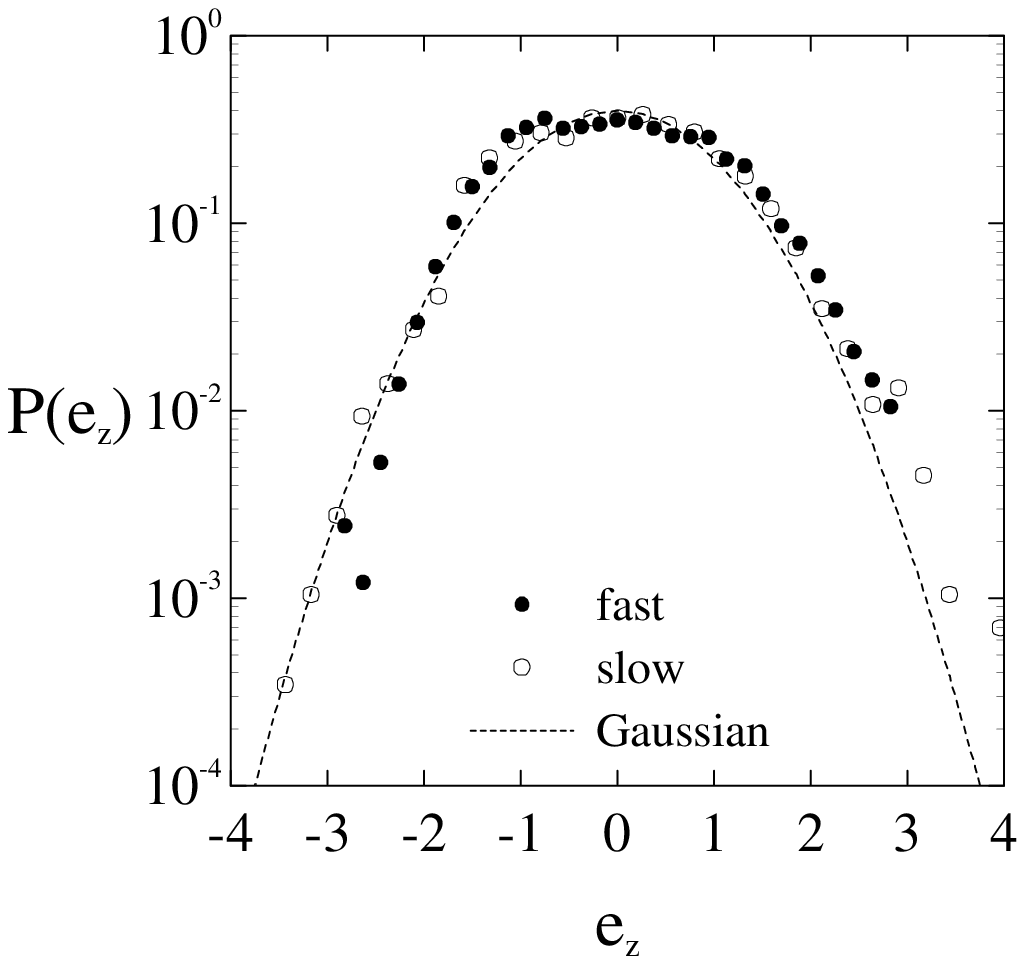}
                     \epsfxsize=6.0 cm \epsfbox{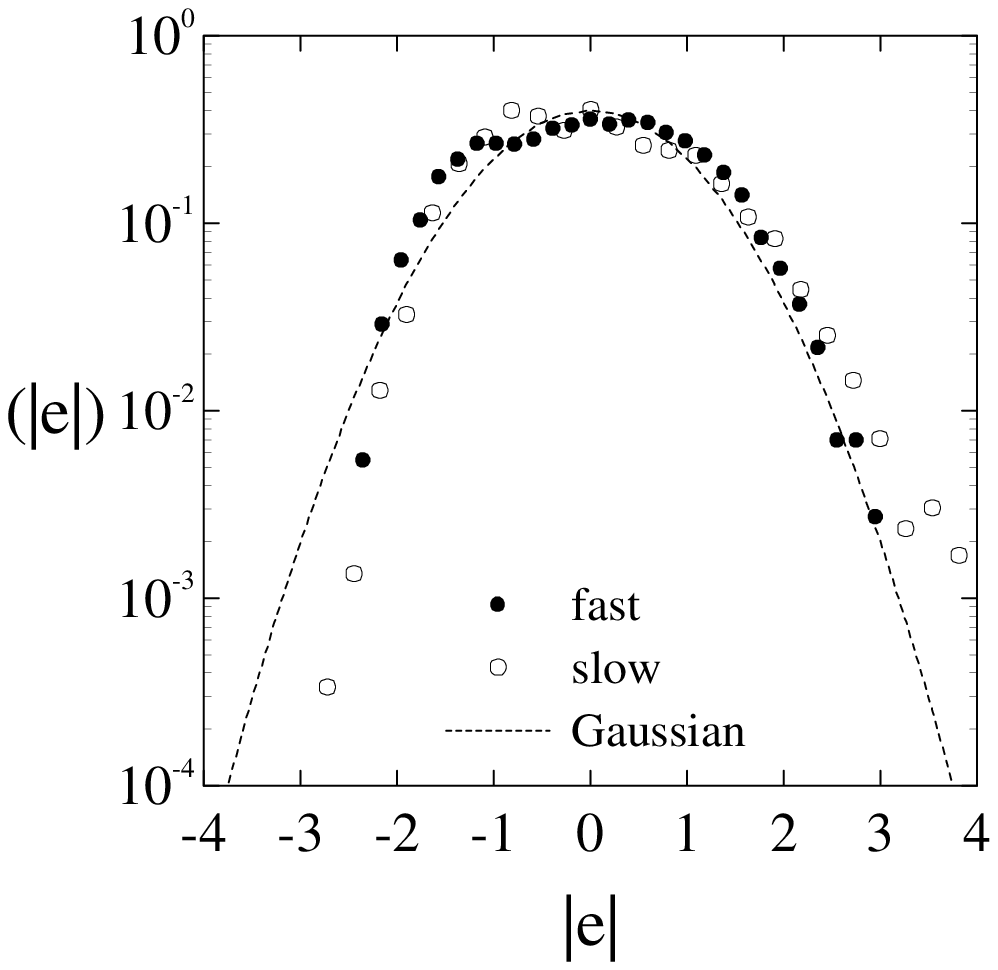}}
                  }
   \caption{The PDFs of induced electric field. From the left column, PDFs refer to $e_v$, 
               $e_b$, $e_y$ and $e_z$ components and the magnitude $|\cdot|$ respectively, 
               as obtasined from $81$ seconds resolution data (see text).
               Solid symbols refer to fast streams, while open symbols refer to slow streams.}
\label{fig:pdfs}
\end{figure}
In figure~\ref{fig:pdfs} we present the PDFs of the electric field. In order to compare
the shape of the PDFs we have previously translated and normalized the variables 
to their standard deviation, so that all PDFs have zero mean and unit standard
deviation. The plots refer to the field components in $V_0$ and $B_0$ frames. 
The PDFs for the $y$ and $z$ components are roughly gaussians for both frames, 
and the same holds for the magnitude $|e|$. 
The $x$ component, aligned either to the radial direction ($e_v$) or to the mean field 
($e_b$), presents a different statistics, depending on the reference frame. In particular, both
in fast and in slow wind, the electric field component along $B_0$ is distributed as a gaussian.
On the other side, the PDFs of the electric field component aligned to $V_0$ show 
exponetial tails, as observed for all components fluctuations in~\cite{breech}.
These observations can be quantitatively measured by computing the flatness 
of the PDFs. The flatness $F(s)=\langle s^4\rangle/\langle s^2\rangle^2$ gives 
informations about the shape of the PDF of a variable $s$. 
In particular, for a gaussian distribution $F=3$. Values of the flatness 
larger than 3 indicate a fatter PDF, namely with higher tails. For $F<3$, conversely, 
the PDF has faster decaying tails. 
The values of the flatness computed from data are in the range $2.4$--$3.4$, except for the electric 
field component along the bulk velocity (radial direction) which has $F=7.1$, for fast and $F=7.7$ for
slow wind. This confirm the direct 
observation of the PDFs. It is worth noting that no difference can be seen here between fast 
and slow streams. However, the separation in fast and slow streams allow for a comparison 
between more homogeneous datasets. We note here that the rotation of the reference frame
do not affect the $y$ and $z$ components, so we only report here the results obtained in the
bulk velocity frame.

In order to compare our datasets with the one used in Ref.~\cite{breech}, we also analysed 
the statistics of the field fluctuations $\gra{e^\prime}$. Our results (not shown) are 
in agreement with~\cite{breech}, namely, we find exponential PDFs for all components, and with 
no remarkable differences between the different reference frames. 

\section{Intermittency}

\begin{figure}
   \centerline{\epsfxsize=8.0 cm \epsfbox{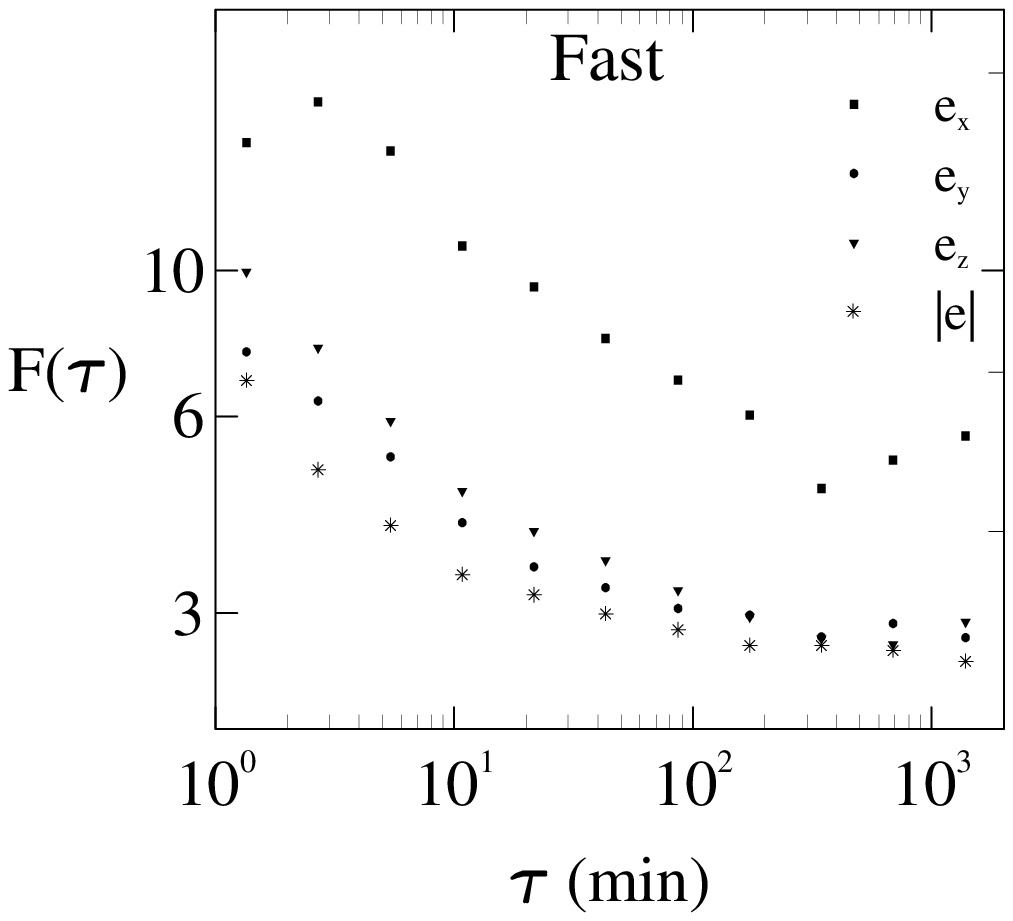}\epsfxsize=8.0 cm \epsfbox{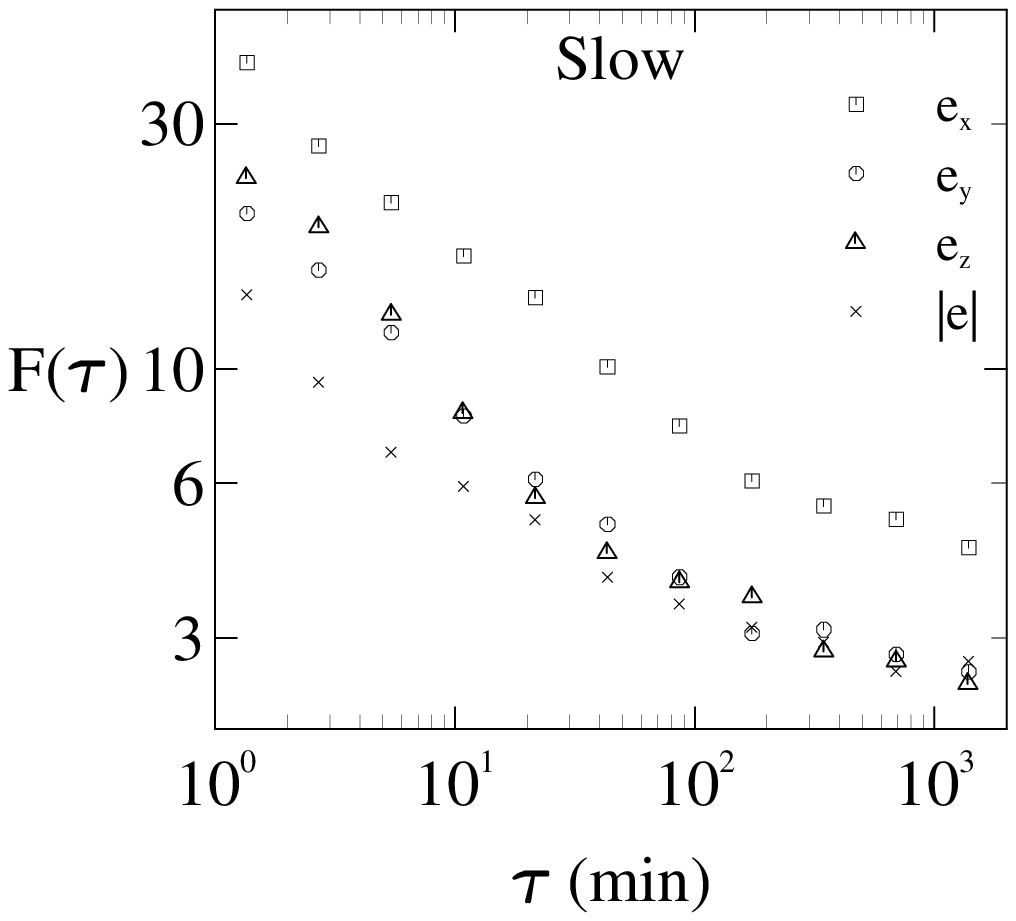}} 
   \caption{The flatness of induced electric field increments for fast streams 
              (letf panel) and slow streams (right panel). Note that the large scale flatness of the 
               $x_v$ component don't reach the gaussian value $F=3$.
               }
\label{fig:flat}
\end{figure}

\begin{figure}
    \centerline{\hbox{\includegraphics[width=10.cm,angle=90]{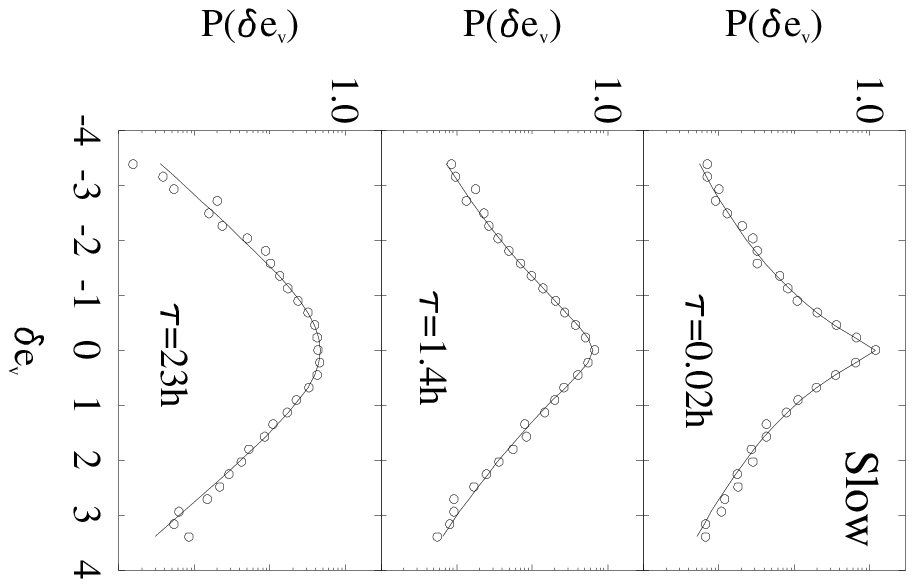}
                             \includegraphics[width=10.cm,angle=90]{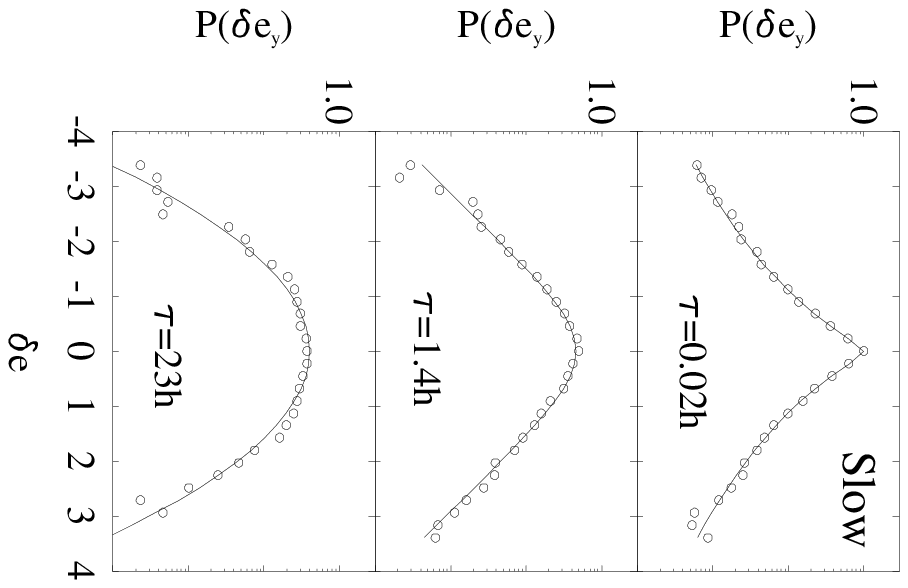}}}
   \caption{The PDFs of induced electric field increments at three different time lags, increasing from top 
               to bottom (indicated), for slow wind, $e_v$ (left) and $e_y$ 
               (right) components. Note the high tails at small scales. The fit with the Castaing PDF (see text)
               is reported as a full line.}
\label{fig:pdf_int}
\end{figure}

\begin{figure}
   \centerline{\epsfxsize=8.0 cm \epsfbox{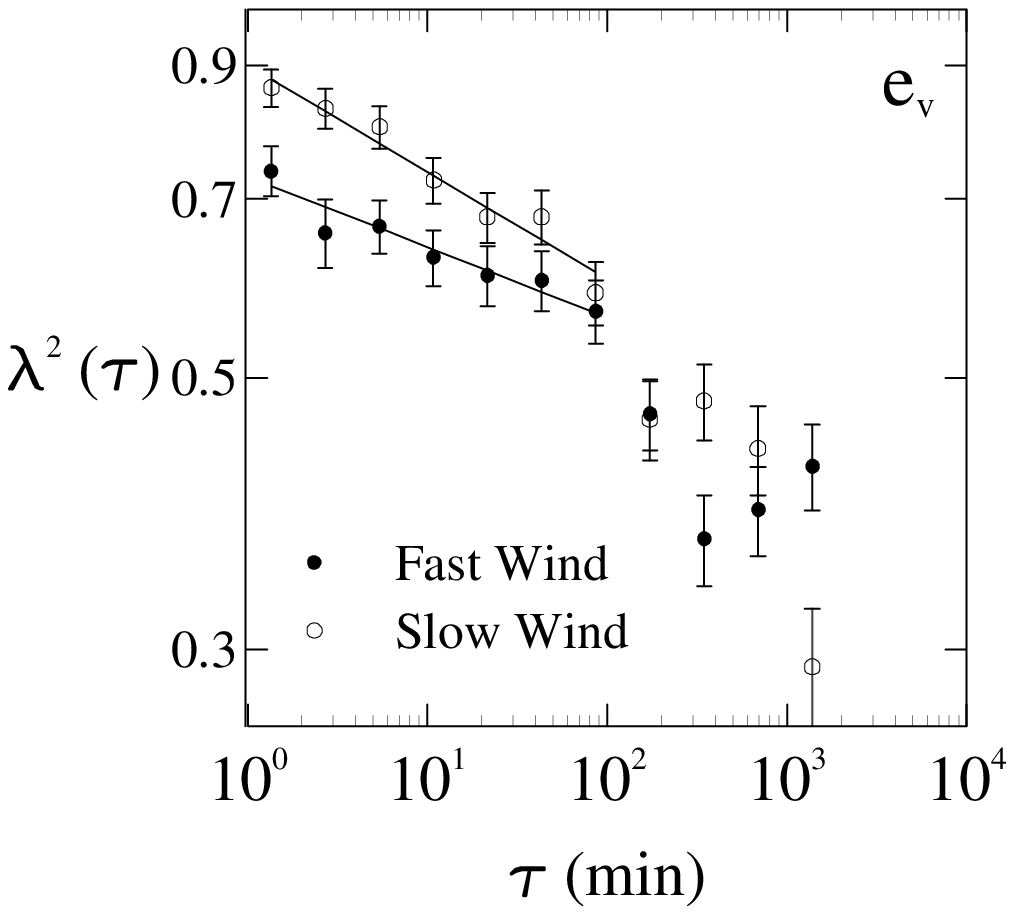}\epsfxsize=8.0 cm \epsfbox{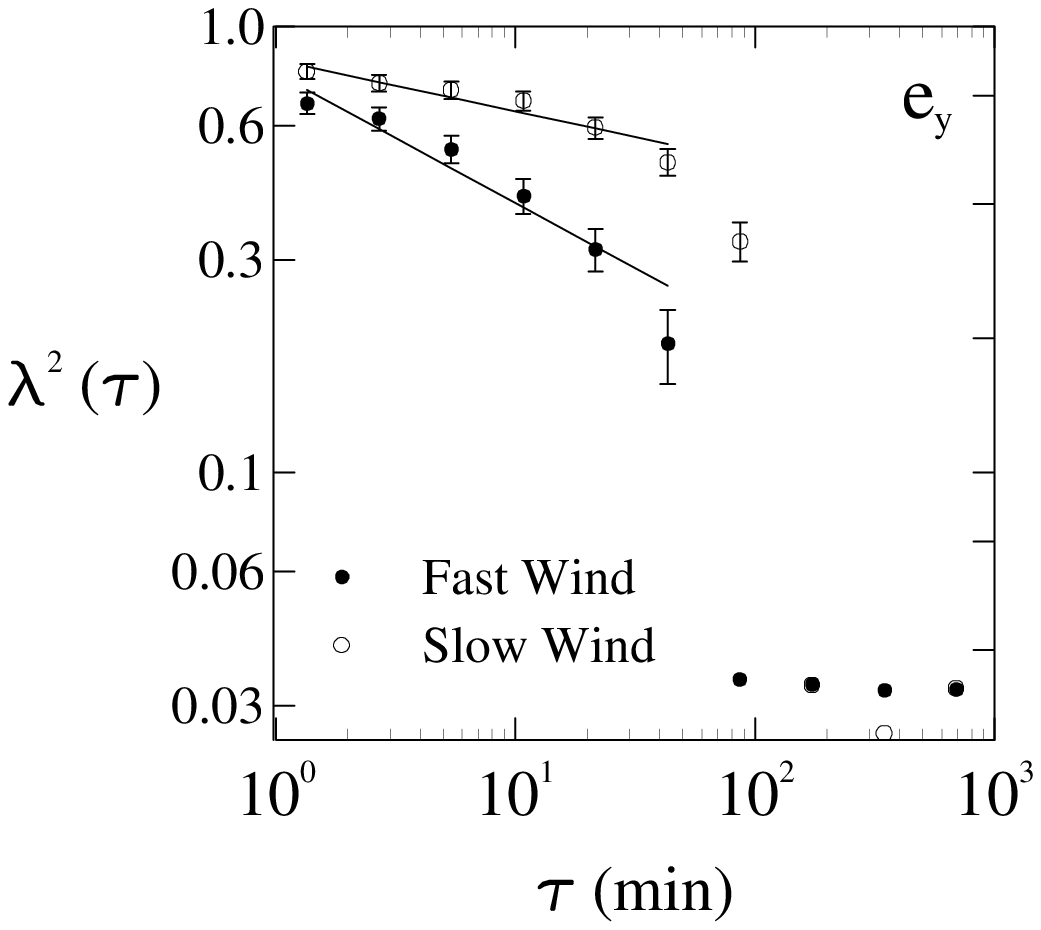}} 
   \caption{The parameter $\lambda^2$ as a function of the time lag $\tau$ as obtained from the fit
               of the measured PDFs with the model discussed in the text.
               Letf panel refers to the $e_v$ component (for both fast and slow streams, as indicated),
               right panel to the $e_y$ component. Fit with power-laws are superimposed as full lines.
               }
\label{fig:lam2}
\end{figure}

In recent years, the solar wind velocity and magnetic field intermittency has been studied in detail
by several autors~\cite{sorriso99,forman}. One suitable approach to intermittency is the 
study of the fields increments~\cite{frisch}, defined as $\delta \psi_\ell (r)=\psi (r+\ell)-\psi(r)$. 
Such variables are used to describe the presence of structures of the field $\psi (r)$ at a given scale $\ell$, 
as for example eddies, shears, strong gradients, shocks and so on. Thus, the statistical properties
of the field increments can give informations about the turbulent energy cascade mechanism, 
responsible for the emergence of structures on a wide range of scales.
In this paper we show that the induced electric field as measured in the solar wind plasma is
characterized by intermittency. To this aim, we compute the PDFs of the increments $\delta e_\tau$, 
where the temporal lags $\tau=\ell/\langle v \rangle$ are used instead of the lenght scales $\ell$ via
the Taylor hypothesis. The intermittency effects can be observed 
as the departure from scaling invariance of the normalized PDFs of the field increments~\cite{vanatta}. 
Figure~\ref{fig:pdf_int} shows the distributions $P(\delta e_\tau)$ for the slow wind, computed for three
different values of the scale. The PDFs clearly display intermittency. 
In order to investigate more quantitatively such behavior, we can analyse the scale dependence of the
flatness (see above). Figure~\ref{fig:flat} reports the values of the flatness of the induced electric field
components and magnitude increments, for both fast and slow wind. As can be seen, the small scale 
increments have high flatness, while the gaussian value $F=3$ is recovered as the scale increases. 
This does not hold for the radial component of the field, for which the asymptotic value of $F$ at large 
scales is considerably higher than for the other components. This is easily understood considering that
the large scale PDF should reproduce the one-point statistics of the field, which is not gaussian
for the $e_v$ component, as we have shown before.
The values of the small scale flatness are also higher for the slow wind than for the fast wind. This 
indicates a higher intermittency in slow wind electric field, and is in accord with previous results
on velocity and magnetic field (see e. g. \cite{sorriso99}). For a better characterization of intermittency,
we use a model for PDF scaling to fit the measured distributions, and then study the scaling properties 
of the parameters obtained from such model~\cite{castaing}.
The model PDF we use here has already been adopted to investigate intermittency in solar wind 
velocity and magnetic field, for example in Refs.~\cite{sorriso99,forman}. Here we shortly describe 
the main idea underlying the model, and we address the reader to the quoted papers for a more 
detailed description. 
In a multifractal picture of turbulence (see for example~\cite{frisch}), 
the PDF of the field increments at a given scale can be 
interpreted as the superposition of many PDFs, each one describing the statistics of the field inside 
a well defined (fractal) subset of the field. Then, the resulting PDF can be computed as the 
sum of such partial PDFs, each one weighted by its relative occurrence in the field. This 
view is supported by the conditional analysis of the solar wind magnetic field 
increments~\cite{sorriso02}. The model PDF can be built up by introducing a parent
distribution, describing the statistics inside each subsets, and a distribution for the weights of the parent
distributions. The parent distribution is the large scale distribution of the field increments, so 
that, as can be checked in Figure~\ref{fig:pdfs}, we choose a gaussian $G_\sigma(\delta e_\tau)$ 
($\sigma$ being the standard deviation). The scale-dependent weights 
PDF is introduced as the distribution $L(\sigma_\tau)$ of the widths $\sigma$, so that the global PDF
can be obtained by computing the convolution of the parent distributions (of variable width $\sigma$) 
with the distribution of their weights. The distribution $L(\sigma_\tau)$ could be in principle observed 
directely from the data, by performing a conditioned analysis. 
Unfortunately, Helios~2 dataset is not large enough to allow for this kind of analysis, 
so that the shape of $L(\sigma_\tau)$ has to be defined by some theoretical arguments. 
As in~\cite{castaing}, we use a Log-normal function 
$L_\lambda(\sigma_\tau)=\exp{(-\frac{\ln^2\sigma}{2\lambda^2}})/\sqrt{4\pi}\lambda$, whose width
$\lambda^2$ determine the shape of the global PDF. In fact for 
$\lambda^2=0$ the Log-normal PDF is a $\delta$-function, so that the convolution gives the 
parent distribution (gaussian). As $\lambda^2$ increases, the convolution includes more and more 
different values of $\sigma$, and with a more and more important weight, so that the tails of the 
resulting PDF raise. Figure~\ref{fig:pdf_int} presents the model PDFs fitted to the data. As can be 
seen the model reproduces in a satisfactory way the scaling evolution of the PDFs.
The scaling properties of the parameter $\lambda^2$ represent an useful tool to 
characterize quantitatively the intermittency~\cite{castaing,sorriso99}. Since in fully developed 
turbulence $\lambda^2\sim \tau^{-\beta}$ display power-law behavior, it turns out 
that the scaling properties of the PDFs, and thus intermittency, can be labeled by a pair of 
parameters, namely: the scaling exponent $\beta$, and the maximum value $\lambda^2_{max}$ 
reached inside the scaling range. 
Figure~\ref{fig:lam2} shows the scaling of the parameter $\lambda^2(\tau)$ obtained from the fit of PDFs,
while in Table~\ref{table:cast} we collect the parameters $\beta$ and $\lambda^2_{max}$ obtained from the
fit with a power-law of $\lambda^2(\tau)$, in a given range of scales.
This has been already done for solar wind velocity and magnetic field~\cite{sorriso_pss}, and we 
report here these results for comparison (see Table~\ref{table:cast}).
As for velocity and magnetic field, the statistical features of the induced electric field are thus well
described by the multifractal model. The scaling properties of $\lambda^2$ are evident and suggest
the presence of an intermittent turbulent cascade of the ideal invariants (energy, magnetic helicity, 
cross-elicity).
From values of the parameter reported in Table~\ref{table:cast}, we can observe that the intermittency 
is more active than for the wind velocity, as in the case of the magnetic field. Higher values of 
$\lambda^2_{max}$, in fact, indicate a higher non-gaussianity of small scale PDFs, which in turn
indicate a larger presence of intermittent structures in the flow. The values of the slope $\beta$ are
small, indicating a slow, not very efficient cascade mechanism. The fact that $x\parallel V_0$ component 
values of $\beta$ are smaller than for the $y$ component is again related to the non-gaussian large 
scale statistics of the field: structures are already present at middle-range scales, and the range of
variation of $\lambda^2(\tau)$ results smaller.

\begin{table}
  \caption{We report the values of the parameters $\lambda^2_{max}$ and $\beta$ 
  obtained from the fitting procedure for $\lambda^2(\tau)$ of the PDF of increments 
  with the multifractal model, for the solar wind induced electric field. 
  For comparison, results for the longitudinal velocity and magnetic field 
  increments are also reported. In all cases, the reference frame is with $x\parallel V_0$}  
  \begin{center} 
  \begin{tabular}{ccc} 
           &  $\lambda^2_{max}$ &  $\beta$  \\ 
  \hline 
  $e_x$ (fast) & $0.74 \pm 0.04$ & $0.06 \pm 0.02$ \\ 
  $e_x$ (slow) & $0.86 \pm 0.03$ & $0.09 \pm 0.01$ \\ 
  $e_y$ (fast) & $0.67 \pm 0.04$ & $0.29 \pm 0.03$ \\ 
  $e_y$ (slow) & $0.79 \pm 0.05$ & $0.12 \pm 0.02$ \\ 
  $b_x$ (fast) & $0.88 \pm 0.04$ & $0.19 \pm 0.02$  \\ 
  $b_x$ (slow) & $0.73 \pm 0.04$ & $0.18 \pm 0.03$ \\ 
  $v_x$ (fast) & $0.51 \pm 0.04$ & $0.44 \pm 0.05$  \\ 
  $v_x$ (slow) & $0.37 \pm 0.03$ & $0.20 \pm 0.04$ \\ 
  [2pt]\end{tabular} 
  \label{table:cast} 
  \end{center} 
\end{table} 
\section{Discussion and conclusions}

We have analysed some statistical properties of the interplanetary induced electric field, 
as measured by Helios~2 spacecraft. The data we used are selected so that our 
samples are homogeneous with respect to velocity and solar activity. 
The one-point statistics of PDFs of the IEF components are gaussian, except for the radial
component, which shows exponential tails. This result may indicate that hypotheses
on correlations, required for the analytical results in~\cite{milano}, do not hold in the solar wind
for the induced electric field. However, exponential PDFs are found found when considering the 
IEF associated with the magnetic field and velocity fluctuations. 
This result is in agreement with Ref.~\cite{breech}). 
We wish to point out that, in the present work, we separate fast from slow wind, since the statistical 
properties of slow and fast wind had been shown to be different~\cite{tuemarsch}.
Moreover, the relatively short-time range covered by the Helios 2 data (about one year) prevent 
from the mixing of different solar activity levels, wich could lead to lack of stationarity of turbulence. 

The analysis of PDFs of the IEF increments at different scales provides the characterization of
intermittency. We performed such analysis using a multifractal model PDF (Castaing distribution) and 
fitting the model to data, in order to obtain the characterizing parameters. 
The scaling properties of such parameters are shown in Figures~\ref{fig:lam2}, that, together with
Figure~\ref{fig:flat}, evidences and quantifies the typical evolution of the statistics, from the large scale 
parent (one-point) distribution, to the high-flatness (high $\lambda^2$), high-tailed small scales statistics. 
A power-law behaviour is found for $\lambda^2$ for two orders of magnitude, in the region between $1$ 
minute and $1$--$2$ hours, that can be view as the inertial range of the turbulent cascading quantities.
This range coincides with previos results obtained from the analysis of velocity and magnetic field 
turbulence (see for example Ref.~\cite{sorriso99}), and confirm one more time the intermittent, turbulent
nature of solar wind fluctuations.
The values of the parameters $\lambda^2_{max}$ and $\beta$, collected in Table~\ref{table:cast}, can 
be interpreted in terms of the topology of the most intermittent structures~\cite{castaing}. This could
then be observed directely from the data, by extracting such structures~\cite{veltri}, and then comparing 
the results. More detailed study of the nature of electric field structures is left for future works.

\acknowledgments
We thank F. Mariani and N. F. Ness, PIs of the magnetic experiment; H. Rosenbauer and R. Schwenn, 
PIs of the plasma experiment onboard Helios 2, for allowing us to use their data.

\end{document}